\documentclass[a4paper]{jpconf}
\usepackage{graphicx}
\usepackage{amsmath}
\usepackage{subcaption}
\usepackage{siunitx}					
\usepackage[colorlinks=true, a4paper=true, pdfstartview=FitV,
linkcolor=black, citecolor=black, urlcolor=black]{hyperref}
\usepackage{url}

\newcommand{\diff}{\mathrm{d}}

\newcommand{\mean}[1]{\langle #1 \rangle}

\usepackage{scalerel}
\newcommand{\sub}[1]{{\scaleto{\mathrm{#1}}{5pt}}}

\DeclareFontFamily{U}{BOONDOX-calo}{\skewchar\font=45 }
\DeclareFontShape{U}{BOONDOX-calo}{m}{n}{
  <-> s*[1.05] BOONDOX-r-calo}{}
\DeclareFontShape{U}{BOONDOX-calo}{b}{n}{
  <-> s*[1.05] BOONDOX-b-calo}{}
\DeclareMathAlphabet{\mathcalboondox}{U}{BOONDOX-calo}{m}{n}
\SetMathAlphabet{\mathcalboondox}{bold}{U}{BOONDOX-calo}{b}{n}
\DeclareMathAlphabet{\mathbcalboondox}{U}{BOONDOX-calo}{b}{n}
\newcommand{\calE}{\mathcalboondox{E}}

\begin{document}

\title{Modeling and simulation of transverse wakefields in PWFA}

\author{J B B Chen$^{1,2}$, D Schulte$^1$ and E Adli$^2$}

\address{$^1$ CERN, CH-1211 Geneva 23, Switzerland}
\address{$^2$ University of Oslo, Oslo, Norway}

\ead{ben.chen@cern.ch}
\ead{daniel.schulte@cern.ch}
\ead{erik.adli@fys.uio.no}

\begin{abstract}
A simplified model describing the PWFA (plasma wakefield acceleration) transverse instability in the form of a wake function parameterized only with an effective cavity aperture radius $a$ is benchmarked against PIC-simulations. This wake function implies a $1/a^4$ scaling of the transverse wakefields, which indicates transverse intra-beam wakefields typically several orders of magnitude higher than in conventional acceleration structures. Furthermore, the wakefield formalism is utilized to perform a parameter study for a \SI{1.5}{\tera\electronvolt} plasma wakefield accelerator, where the constraint on drive beam to main beam efficiency imposed by transverse wakefields is taken into account. Eventually, a parameter set with promising properties in terms of energy spread, stability and luminosity per power was found.

\end{abstract}

\section{Introduction}
PWFA is one of the most promising novel acceleration technologies able to generate accelerating gradients in the multi-\SI{}{\giga\volt/\meter} level \cite{Blumenfeld}. There are however challenges that still need to be addressed before this technology can be applied to a future linear collider. Transverse instabilities caused by transverse wakefields, which are fields generated by a driving particle's interaction with the accelerating cavity due to misalignment, are considered one of the main challenges, as this is known to constrain the drive beam to main beam efficiency in CLIC \cite{CLIC_CDR}. Transverse wakefields in PWFA can be several orders of magnitude larger than in metallic cavities due to the significantly smaller dimension of a plasma ion bubble, so a good understanding of possible mitigation methods is therefore necessary for a global parameter optimization for a PWFA-LC (plasma wakefield acceleration linear collider). One such mitigation method is BNS damping \cite{BNS}, a well-known technique in RF accelerators, where a correlated energy spread is induced along the beam to disrupt the coherence buildup of transverse oscillations.

Several conceptual parameter sets for a PWFA-LC have been proposed to identify the main challenges and base parameters, one example being the Snowmass parameter set \cite{Snowmass_Erik}. However, in contrast to CLIC, the effect of transverse wakefields on efficiency has so far not been taken into account in PWFA-LC parameter studies, even though Lebedev et al. have studied the relationship between efficiency and instability, and derived an analytical expression \cite{Lebedev_2017}. In this paper, we will conduct a parameter study of the efficiency of a \SI{1.5}{\tera\electronvolt} plasma wakefield accelerator using the Snowmass parameter set as a basis, but taking into account transverse wakefield and the damping effect of energy spread using the approach of a parameter scan.

\section{Transverse wake function}
Plasma acceleration is very computationally expensive to simulate, hence it is very challenging to consider the effect of transverse instabilities on efficiency using traditional PIC-codes. Several studies have proposed simplified models for transverse beam motion in PWFA using coupled differential equations for the beam and plasma channel centroid \cite{Huang, Mehrling1, Mehrling2}. In this paper, we assume that the transverse forces can be expressed using the wake function formalism \cite{wakefield_Wilson,wakefield_Chao}, which is used for describing the well-known BBU-instability in RF accelerators, and will allow for easier comparison with RF accelerators.

In CLIC \cite{RAST_Daniel}, single beam transverse wakefield for small distances between a driving particle located at $\xi'$ and a witness particle located at $\xi$ is modelled using \begin{equation}
    W_\perp(\xi'-\xi)=\frac{2}{\pi\varepsilon_0}\frac{\xi'-\xi}{a^4}\Theta(\xi'-\xi),
\label{eq:transWakeFunction}
\end{equation}
where $\varepsilon_0$ is the permittivity in vacuum, $a$ is the accelerating structure iris radius and $\Theta(\xi)$ is the Heaviside step function. The structure iris is however not well-defined for a plasma, but an effective structure iris \cite{Lebedev_2017, G.Stupakov} can be defined by $a=r_\mathrm{b}(\xi')+\alpha k_\mathrm{p}^{-1}$. Here $r_\mathrm{b}(\xi')$ is the plasma bubble radius at the location of the driving particle, $\alpha$ a numerical coefficient on the order of one, and the plasma skin depth $k_\mathrm{p}^{-1}$ accounts for the penetration depth of the electromagnetic fields. Equation \eqref{eq:transWakeFunction} along with the modification $a=r_\mathrm{b}(\xi')+\alpha k_\mathrm{p}^{-1}$ has been proposed for the PWFA blowout regime in \cite{G.Stupakov}. In this paper, we adopt this wake function, and use the value $\alpha=0.75$, which is the same value used in \cite{G.Stupakov}.

For a beam slice with charge $q$ located at $\xi$, the transverse wake force per unit charge is given by a convolution integral  
\begin{equation}
    \frac{F_\perp(\xi,s)}{q} =  -e\int\limits_{\xi_\sub{H}}^\xi\! W_\perp(\xi'-\xi) \lambda(\xi') X(\xi',s) \,\diff\xi',
\label{eq:wakeForce}
\end{equation}
where $e$ is the elementary charge, $\xi_\sub{H}$ is the longitudinal position of the beam head, $\lambda(\xi)$ is the longitudinal number density of the main beam and $X(\xi,s)$ is the mean transverse offset of the beam slice located at $\xi$.

\begin{figure}[h]
        \includegraphics[width=15pc]{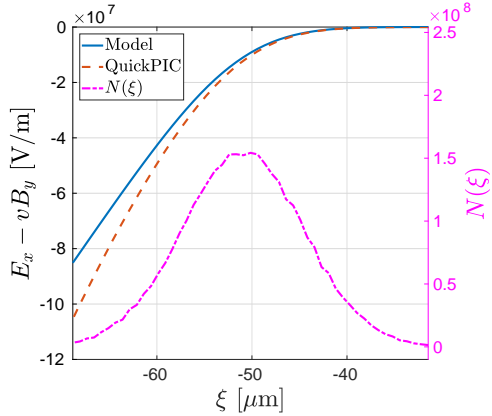}\hspace{2pc}%
        \begin{minipage}[b]{16pc}\caption{The initial transverse force per unit witness charge $F_\perp(\xi,s=0)$ for a main beam with constant transverse offset calculated with equation \eqref{eq:wakeForce} and calculated directly from the output fields of a QuickPIC simulation measured on axis. The longitudinal particle distribution $N(\xi)$ of the main beam is also included in the figure. The beam propagates towards higher values of $\xi$.}
        \label{fig:2019-10-17_wake_convolution_Stupakov-model_s=0m}
    \end{minipage}
\end{figure}

Equation \eqref{eq:wakeForce} gives the transverse force along the main beam after a propagation length $s$, as is illustrated in figure \ref{fig:2019-10-17_wake_convolution_Stupakov-model_s=0m} for a main beam with constant transverse offset propagating along the $\xi$-axis.

\begin{figure}[ht]
    \centering
    \begin{subfigure}[t]{0.32\textwidth}
        \centering
        \includegraphics[width=\columnwidth]{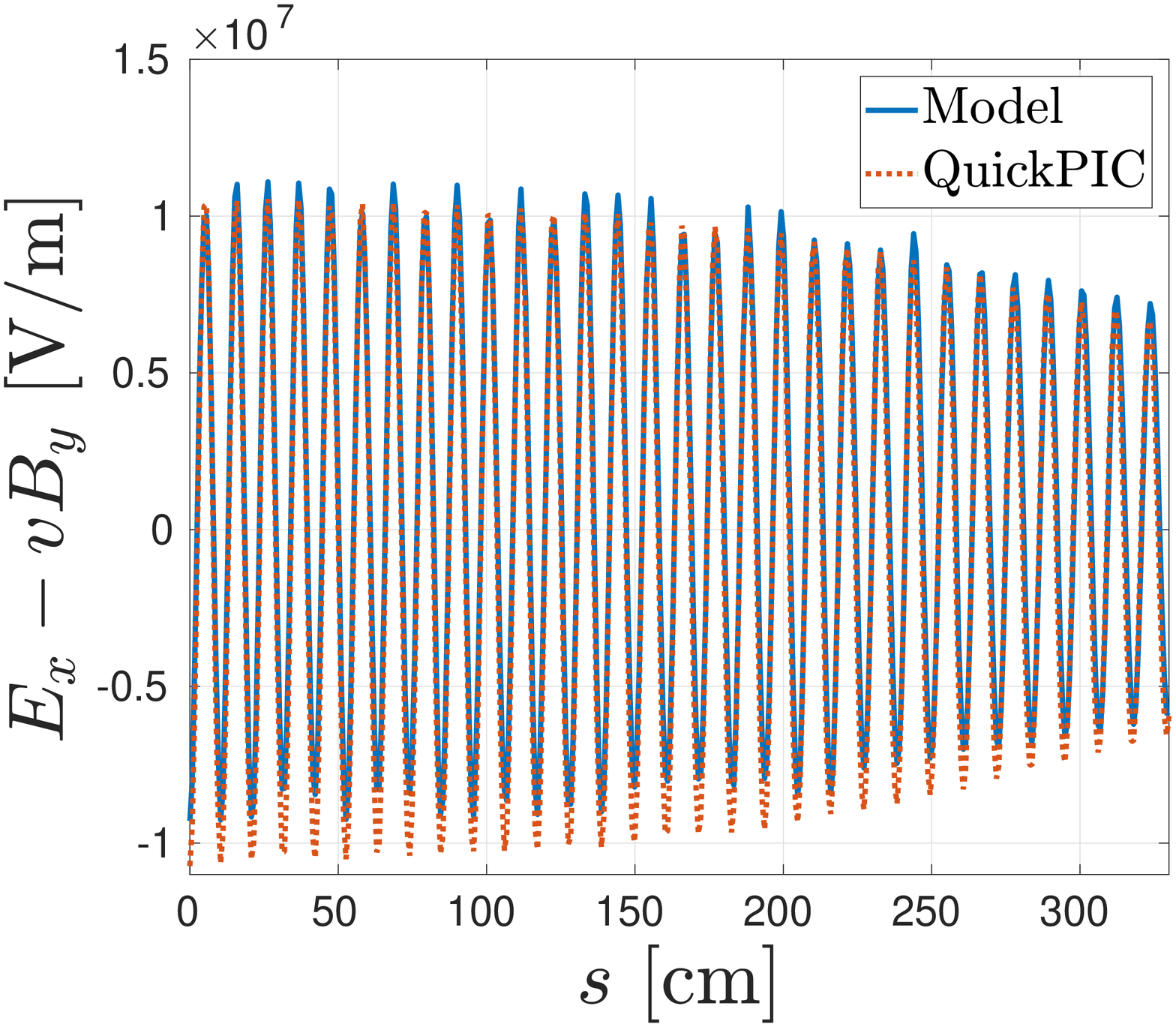}
        \caption{Beam center.}
        \label{fig:transWakeEvolution_0sigmaZ_from_center_slice_Stupakov}
    \end{subfigure}\hspace{0.35pc}
    \begin{subfigure}[t]{0.32\textwidth}
        \centering
        \includegraphics[width=\columnwidth]{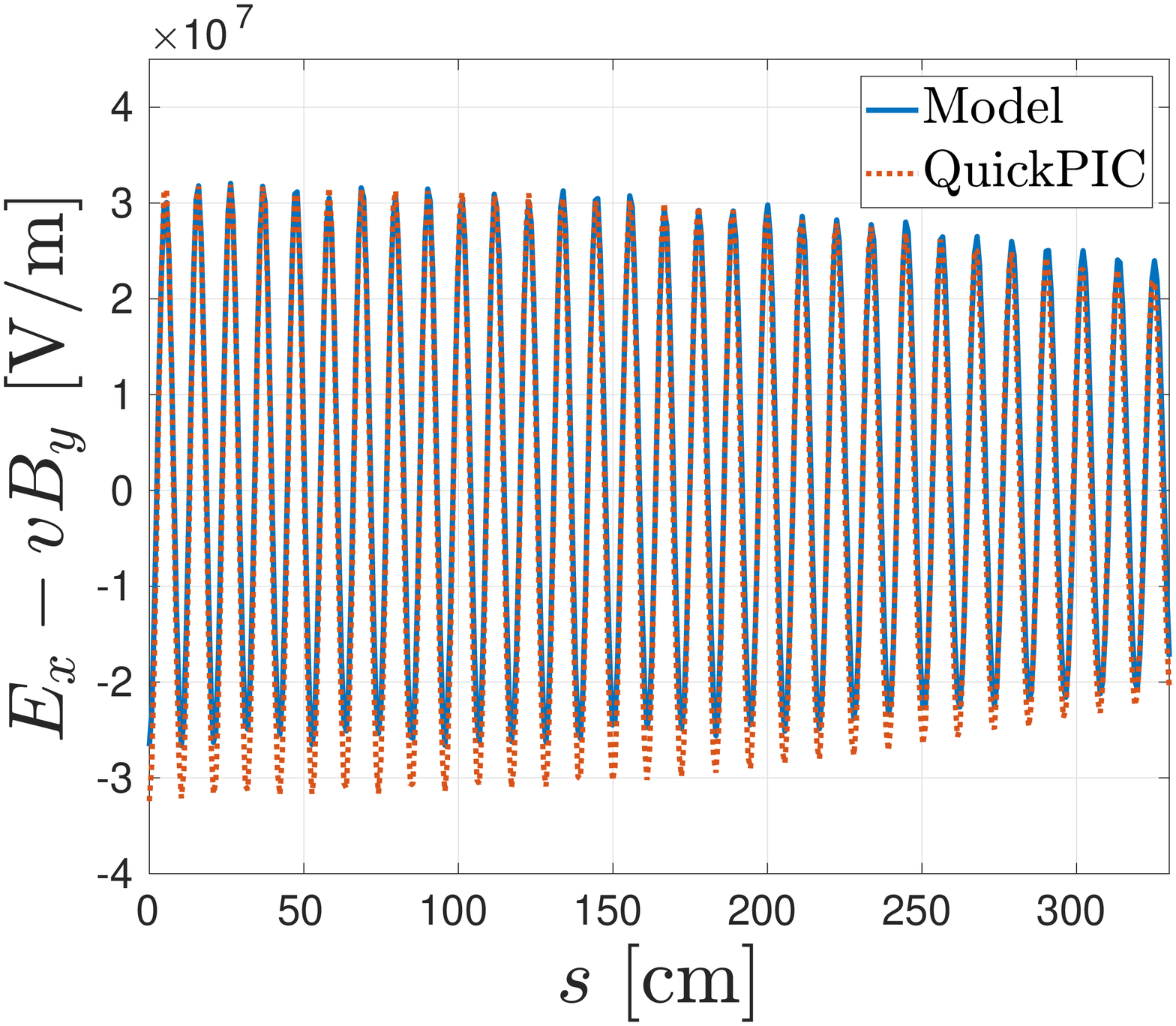}
        \caption{One $\sigma_z$ behind the beam center.}
    \end{subfigure}\hspace{0.35pc}
    \begin{subfigure}[t]{0.32\textwidth}
        \centering
        \includegraphics[width=\columnwidth]{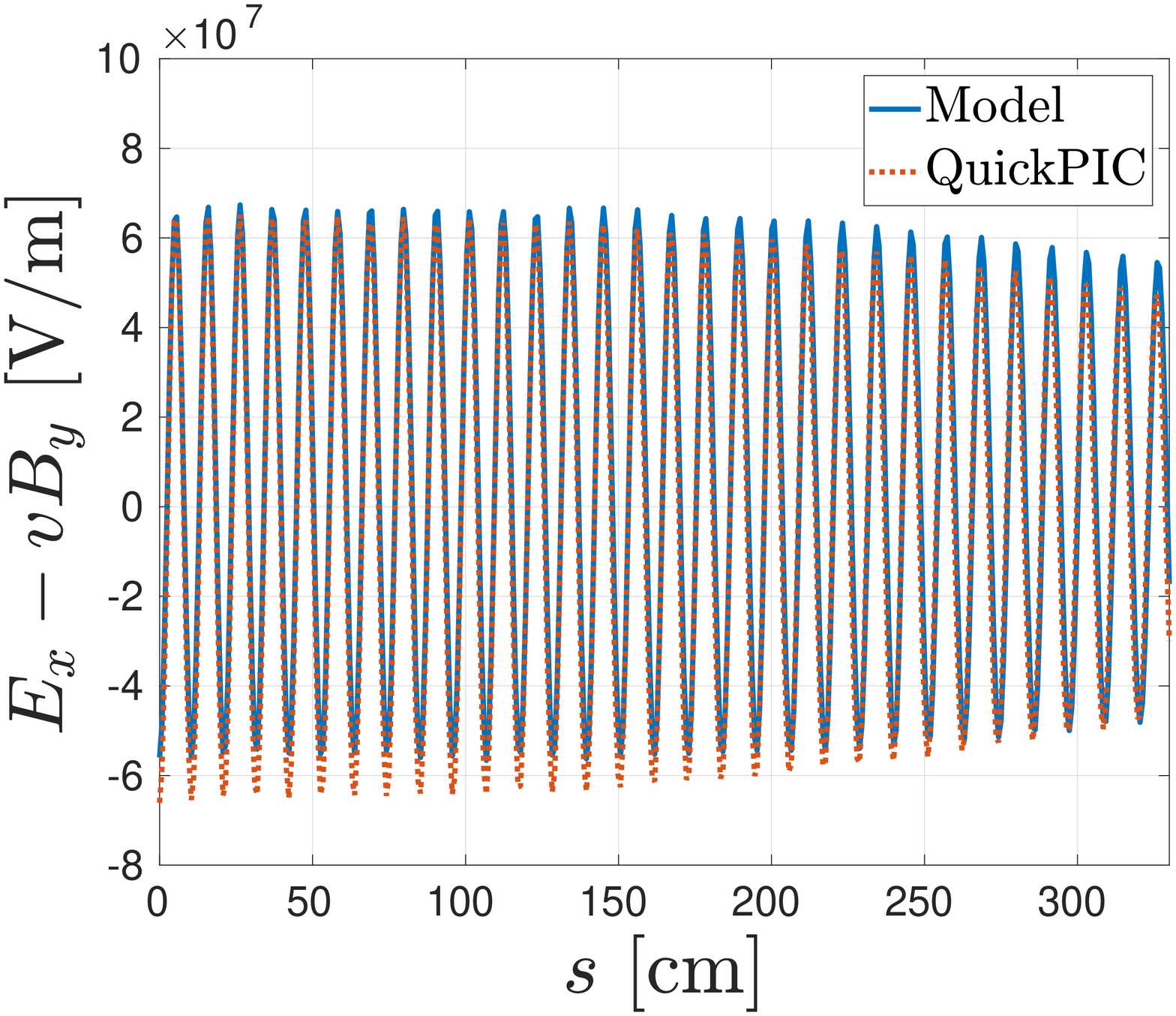}
        \caption{$2\sigma_z$ behind the beam center.}
        \label{fig:transWakeEvolution_-2sigmaZ_from_center_slice_Stupakov}
    \end{subfigure}
    
\caption{Evolution of the transverse wake force per unit charge at main beam slices at various positions along the beam. The fields extracted from QuickPIC are measured on axis.}
\end{figure}

The evolution of the transverse force at three different beam slices is benchmarked against QuickPIC \cite{QuickPIC_Huang} simulation results, where $\lambda(\xi)$ and $X(\xi)$ in equation \eqref{eq:wakeForce} are extracted from QuickPIC simulations, and the transverse force predicted by the convolution integral is then compared against the corresponding fields extracted from QuickPIC results. Figure \ref{fig:transWakeEvolution_0sigmaZ_from_center_slice_Stupakov}-\ref{fig:transWakeEvolution_-2sigmaZ_from_center_slice_Stupakov} show the evolution of the transverse wake on beam slices located 0-2 $\sigma_z$ behind the main beam center. Inside the ion bubble, the transverse fields acting on the main beam consist of the background ion focusing and intra-beam wakefields, which are similar to dipole fields. To avoid noise, the dipole fields extracted from QuickPIC are measured on axis. Except for the small disagreement for negative amplitudes, the model shows a good agreement with simulations.









\section{Simplified quasi-static model}
A simplified quasi-static \cite{QuickPIC_Huang} model was developed in order to exploit the wake function formalism to provide an efficient way of studying transverse instabilities in the main beam in the blow-out regime of a \SI{}{\tera\electronvolt}-scale plasma collider. Similarly to QuickPIC \cite{QuickPIC_Huang}, the simplified model also utilizes the quasi-static approximation, where it is assumed that the ultra-relativistic beam evolves over a much longer time scale compared to the plasma. Mathematically, this is described by the coordinate transformation $(x,y,z,t)\to (x,y,\xi=z-ct,s=ct)$. The time derivative can then be written as 
\begin{equation}
    \frac{\partial}{\partial t} = \frac{\partial\xi}{\partial t}\frac{\partial}{\partial\xi} + \frac{\partial s}{\partial t}\frac{\partial}{\partial s} =  -c\frac{\partial}{\partial\xi} + c\frac{\partial}{\partial s}.
\end{equation}
For an ultra-relativistic beam particle, $\partial_s \gg \partial_\xi$ so that $\partial_t\approx c\partial_s$.

The main beam is sliced longitudinally into slices with equal thicknesses. Assuming that the main beam placed inside the plasma ion bubble does not penetrate the plasma bubble boundary during propagation, we then make the ansatz that the the transverse oscillation of a beam slice located at $\xi$ can be described by 
\begin{equation}
    \frac{\partial^2}{\partial s^2}X(\xi,s) + \frac{1}{\beta(\xi,s)^2}X(\xi,s) = \frac{e^2}{\calE(\xi,s)}\mathcal{W}_\perp(\xi,s),
\label{eq:transverseOscillation}
\end{equation}
where $\beta(\xi,s)=\sqrt{2\gamma(\xi,s)}/k_\mathrm{p}$ is the beta function, and $\calE(\xi,s)=\gamma(\xi,s)m_\mathrm{e}c^2$ is the electron energy of an electron located at $\xi$, that has been accelerated by the longitudinal field $E_z(\xi)$ for a distance $s$. The second term of equation \eqref{eq:transverseOscillation} represents the betatron oscillation caused by the focusing forces of the ion background, while the driving term is attributed to the transverse wakefields. All the preceding slices contribute to the driving term through the convolution integral
\begin{equation}
    \mathcal{W}_\perp(\xi,s) = \int\limits_{\xi_\sub{H}}^\xi\! W_\perp(\xi'-\xi) \lambda(\xi') X(\xi',s) \,\diff\xi'.
\label{eq:transWake}
\end{equation}


The interaction with the plasma and drive beam is represented by the $1/a^4$-dependence of the wake function, and the interaction with the total longitudinal wakefield $E_\parallel(\xi)$. $r_\mathrm{b}(\xi)$ and $E_\parallel(\xi)$ are however not described by this model, and was calculated numerically with QuickPIC in this study. Assuming that $r_\mathrm{b}(\xi)$ and $E_\parallel(\xi)$ do not change significantly during propagation, these quantities only needed to be calculated once in QuickPIC.

These equations are then solved numerically with the quasi-static approximation where the main beam is evolved in $s$, alternating between propagation with frozen transverse forces and interaction with the plasma ion bubble through equation \eqref{eq:transWake} and \eqref{eq:transWakeFunction}, where the transverse forces are updated.

\begin{figure}[ht]
    \centering
    \begin{subfigure}[t]{0.32\textwidth}
        \centering
        \includegraphics[width=\columnwidth]{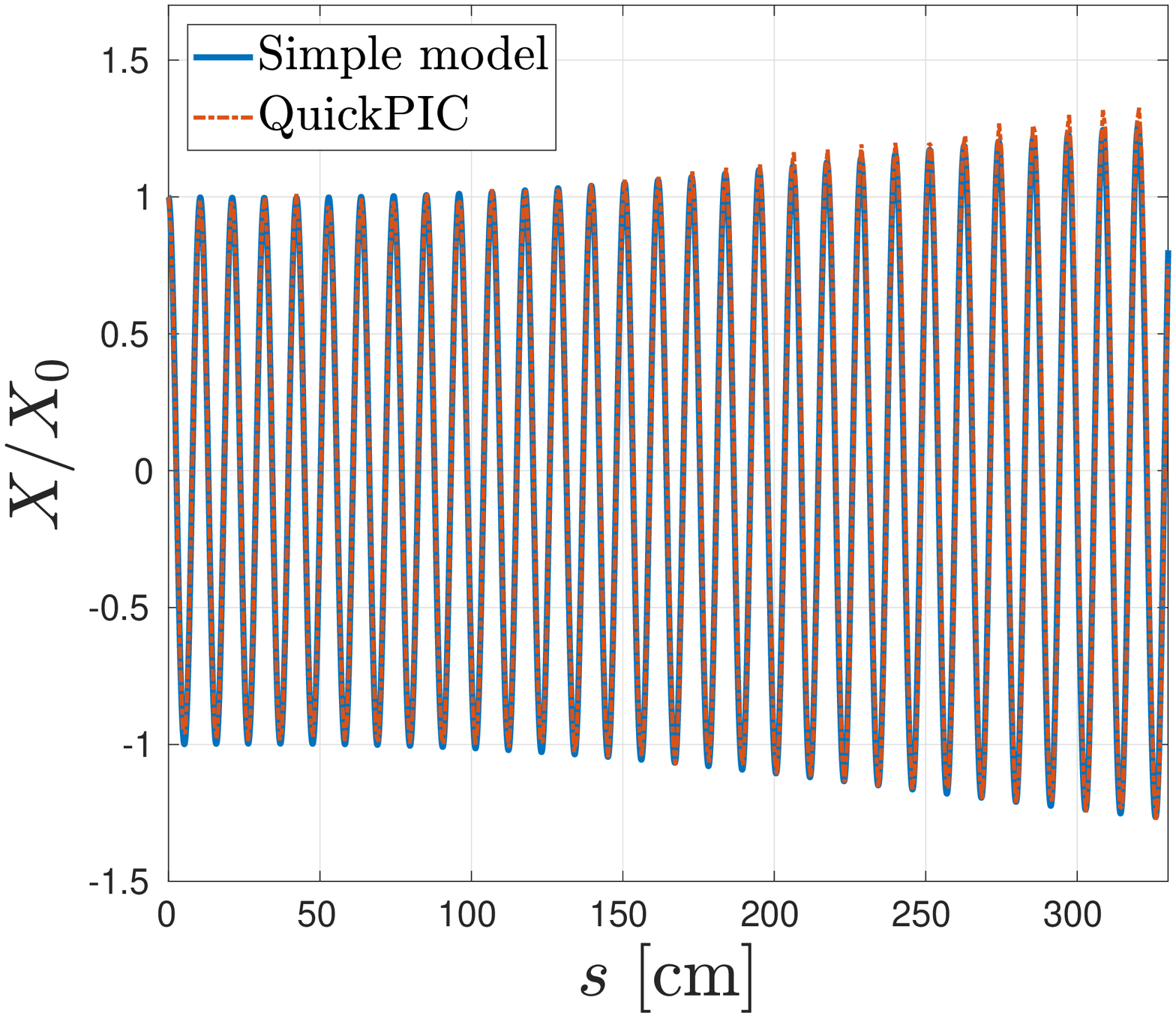}
        \caption{Beam center.}
        \label{fig:2019-03-20_meanX_0sigmaZ_slice_Stupakov}
    \end{subfigure}\hspace{0.35pc}
    \begin{subfigure}[t]{0.32\textwidth}
        \centering
        \includegraphics[width=\columnwidth]{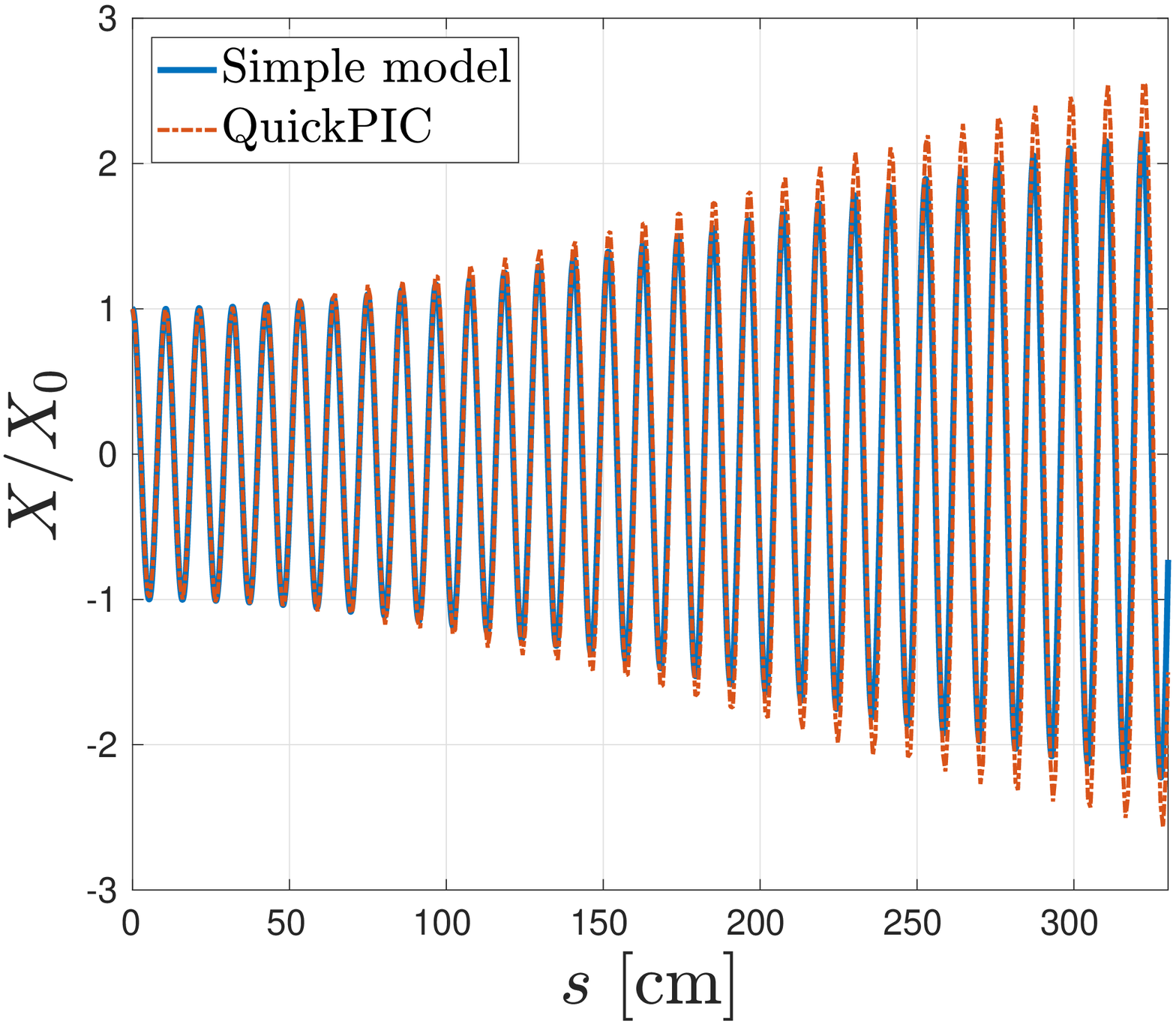}
        \caption{One $\sigma_z$ behind the beam center.}
    \end{subfigure}\hspace{0.35pc}
    \begin{subfigure}[t]{0.32\textwidth}
        \centering
        \includegraphics[width=\columnwidth]{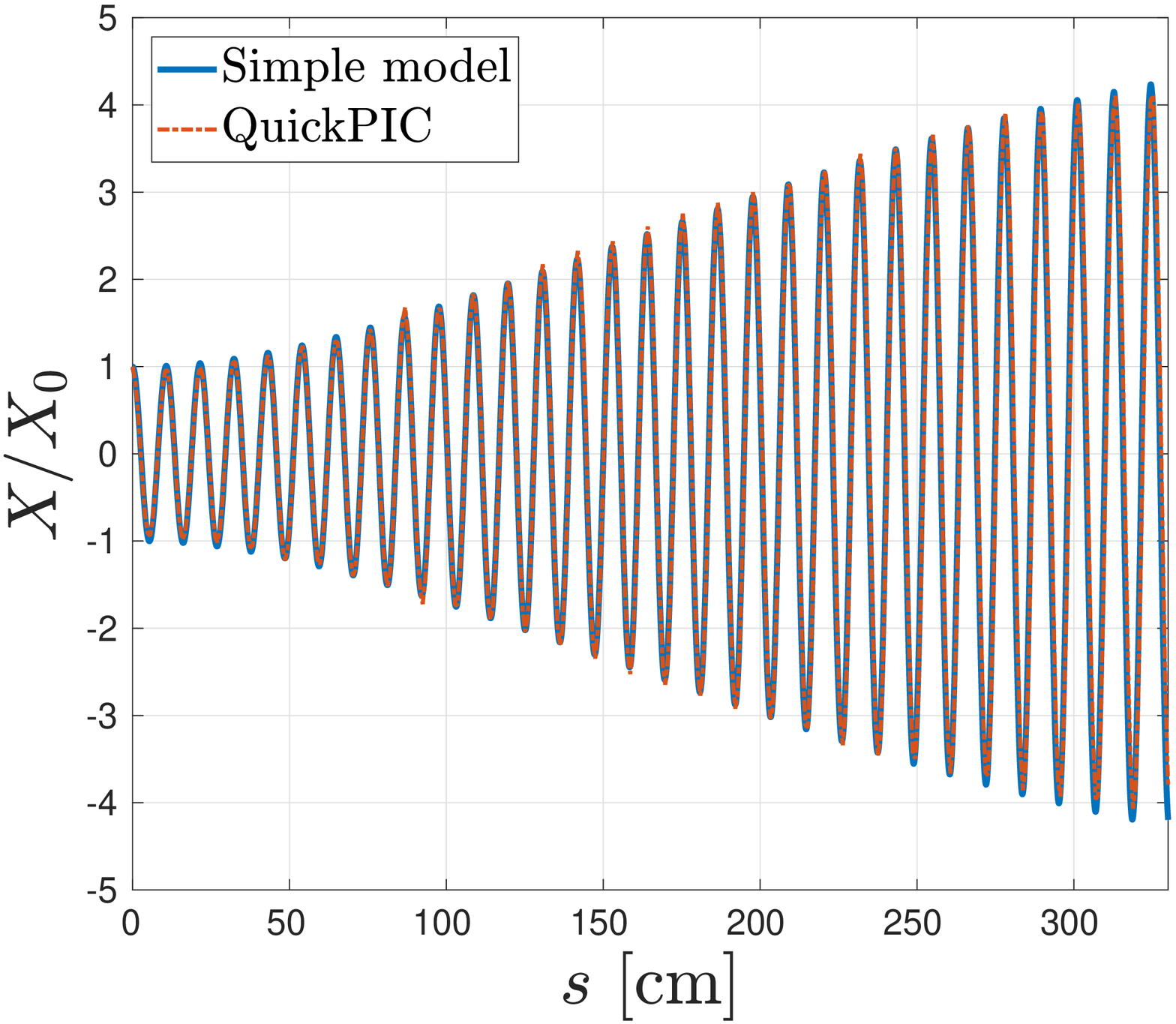}
        \caption{$2\sigma_z$ behind the beam center.}
        \label{fig:2019-03-20_meanX_-2sigmaZ_slice_Stupakov} 
    \end{subfigure}
    
\caption{Comparison of the mean transverse position of main beam slices located at various positions. $X_0=\SI{3.65}{\micro\meter}$.}
\end{figure}

This model was benchmarked against QuickPIC by comparing the mean transverse offset of beam slices located 0-2 $\sigma_z$ behind the beam center. The results are shown in figure \ref{fig:2019-03-20_meanX_0sigmaZ_slice_Stupakov}-\ref{fig:2019-03-20_meanX_-2sigmaZ_slice_Stupakov}. The simplified model agrees very well with the simulation results as long as the main assumptions are valid. The initial offset $X_0$ was chosen to be $X_0=\SI{3.65}{\micro\meter}$, which is on the order of one $\sigma_x$.








\section{Evaluation of the Snowmass parameter set}

\begin{figure}[ht]
\begin{minipage}[t]{18pc}

    \centering
    \includegraphics[width=0.75\columnwidth]{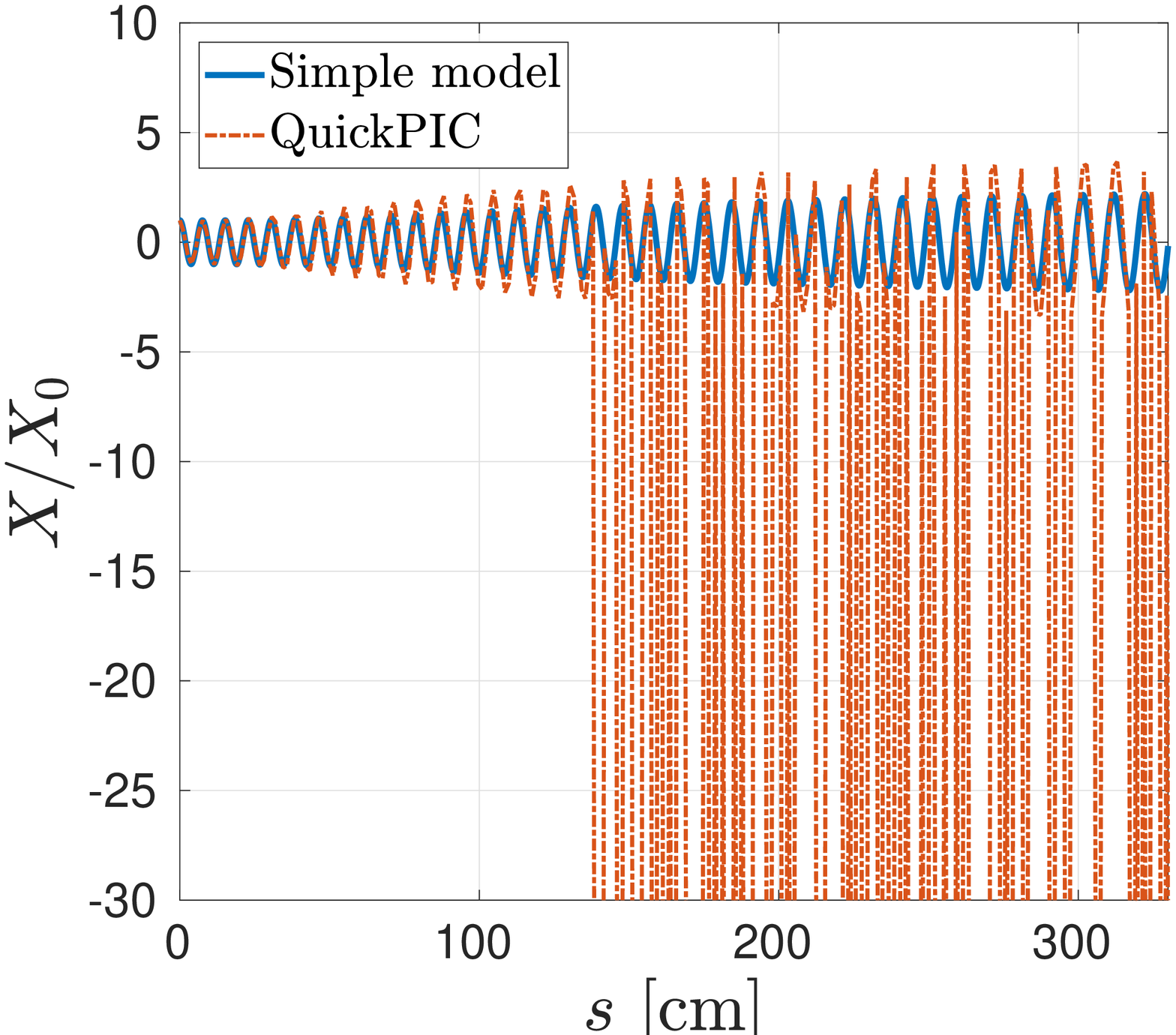}
    \caption{Comparison of the mean transverse position of the main beam slice located at the center of the beam. The Snowmass parameters were used in this simulation, and resulted in a highly unstable main beam after the beam tail came into contact with the bubble boundary, as seen in figure \ref{fig:2019-06-28_QEB+QEP}. This transverse motion can thus not be described with equation \eqref{eq:transverseOscillation}.}
    \label{fig:2019-06-28_meanX_0sigmaZ_slice_Stupakov}

\end{minipage}\hspace{1pc}%
\begin{minipage}[t]{20pc}

    \centering
    \includegraphics[width=0.7\columnwidth]{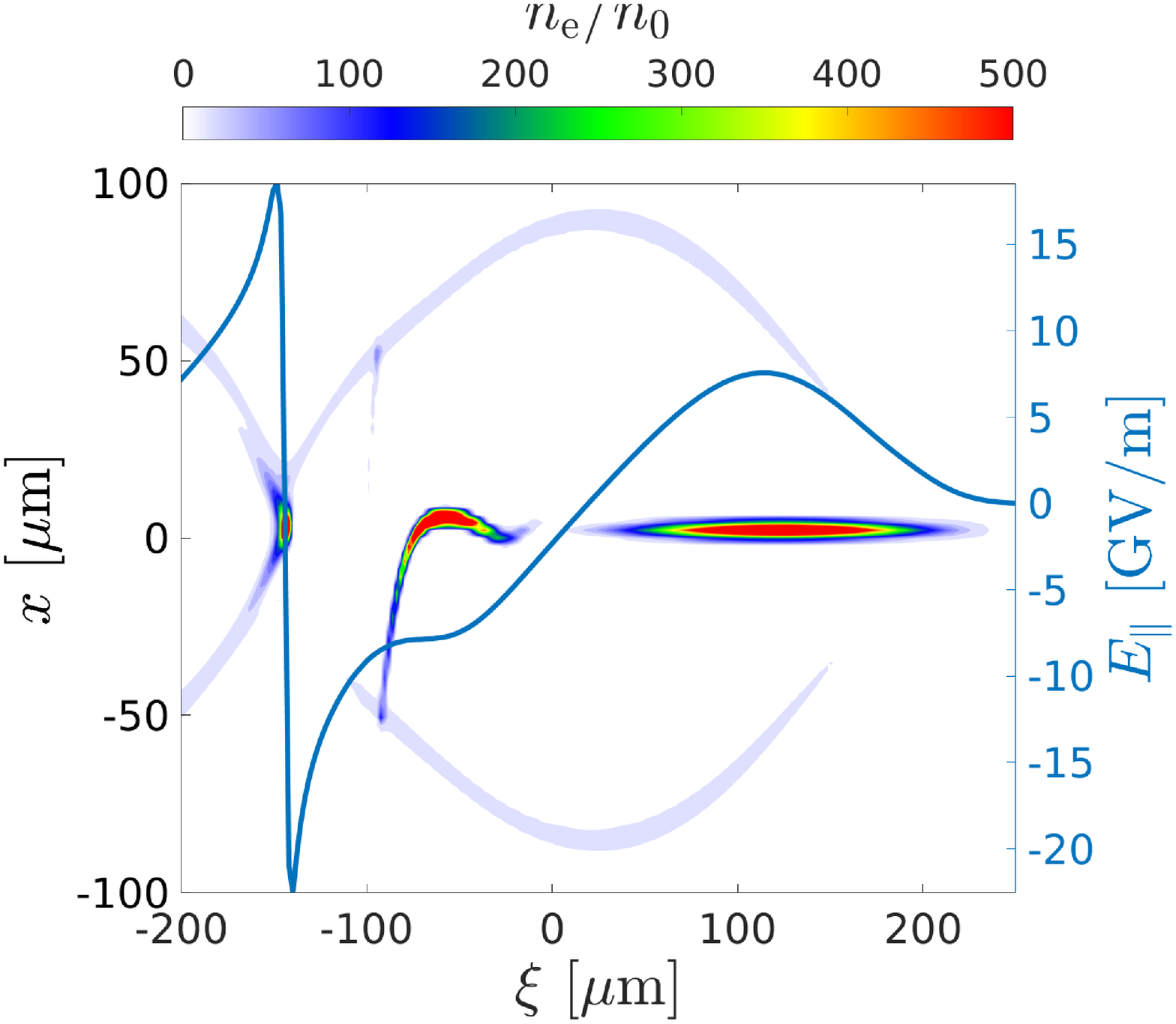}
    \caption{Electron number density $n_\mathrm{e}$ per unit initial plasma density $n_0$ and the total longitudinal electric field $E_\parallel(\xi)$ for $s\approx\SI{140}{\centi\meter}$ obtained from QuickPIC simulation with Snowmass parameters. The plasma electron density has been increased by a factor 10 in order to highlight the bubble boundary.}
    \label{fig:2019-06-28_QEB+QEP}
    
\end{minipage}\hspace{1pc}
\end{figure}

Figure \ref{fig:2019-06-28_meanX_0sigmaZ_slice_Stupakov} compares a results from a QuickPIC simulation against the simplified model using the Snowmass parameter set. The QuickPIC results show that the Snowmass parameter set produced a highly unstable main beam, and eventually caused the tail of the main beam to come into contact with the bubble boundary. It can be seen that the simplified model and QuickPIC were in good agreement until the beam tail penetrated the plasma at $s\approx\SI{140}{\centi\meter}$, as is depicted in figure \ref{fig:2019-06-28_QEB+QEP}, after which the transverse motion of the beam could not be described by equation \eqref{eq:transverseOscillation}. Such unstable cases are however irrelevant for this study, as this paper aims to find a set of parameters for a stable main beam, and not to model highly unstable oscillations.

Nonetheless, because of the unstable beam, the Snowmass parameter set has to be modified in order to achieve stable propagation with high efficiency and low energy spread. This is done in section \ref{sec:parameterStudy}, where we conduct a parameter study of a \SI{1.5}{\tera\electronvolt} plasma wakefield accelerator using the Snowmass parameter set as a basis. 

\section{Parameter study for a \SI{1.5}{\tera\electronvolt} plasma wakefield acclerator}
\label{sec:parameterStudy}

\subsection{Energy spread, instability and efficiency}
\begin{figure}[ht]
    \centering
    \begin{subfigure}[t]{0.32\textwidth}
        \centering
        \includegraphics[width=\columnwidth]{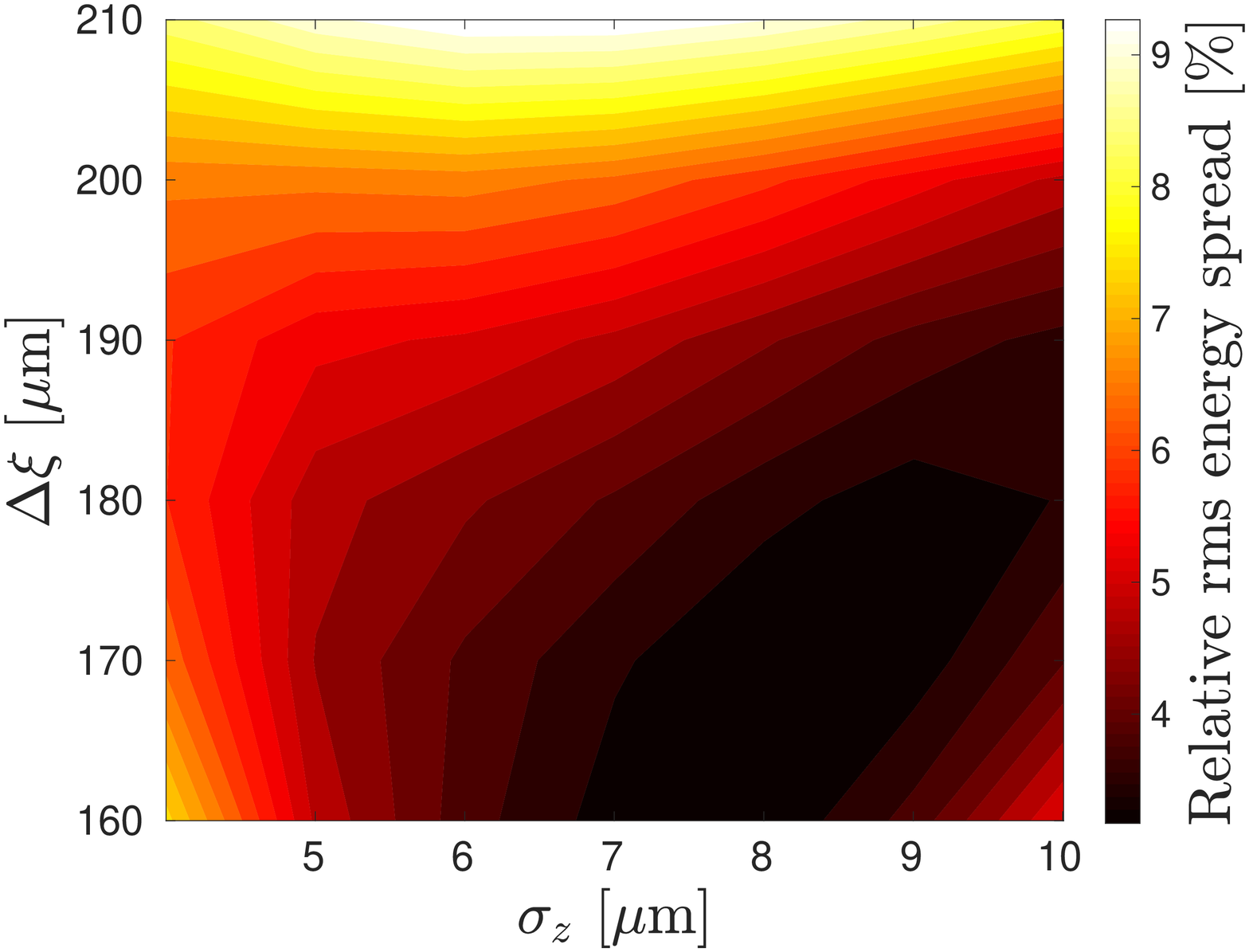}
        \caption{$N_\mathrm{MB}=\SI{8e9}{}$.}
        \label{fig:2019-07-19_sigmaE_E_inf_contourPlot_N8e+09}
    \end{subfigure}\hspace{0.35pc}
    \begin{subfigure}[t]{0.32\textwidth}
        \centering
        \includegraphics[width=\columnwidth]{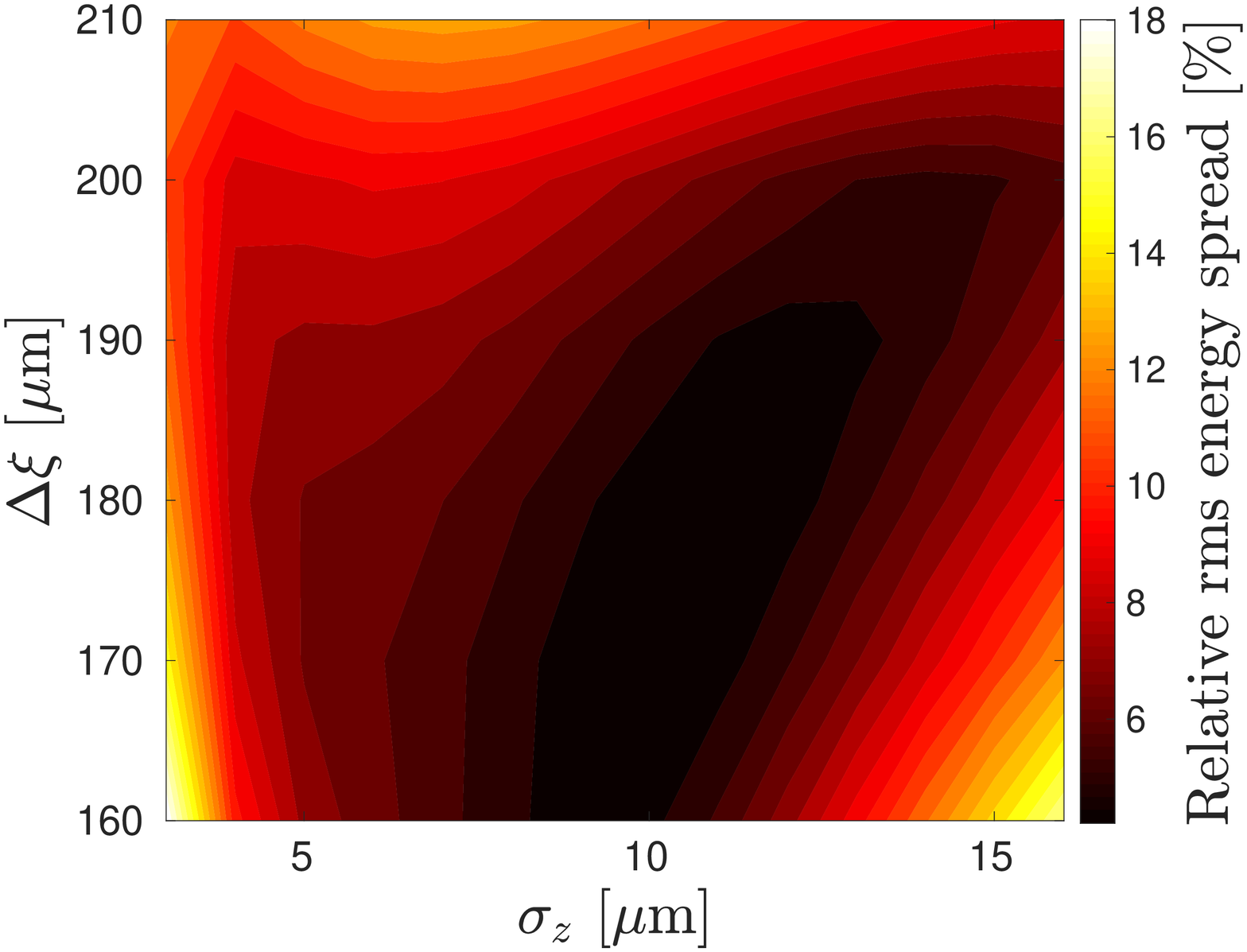}
        \caption{$N_\mathrm{MB}=\SI{9e9}{}$.}
    \end{subfigure}\hspace{0.35pc}
    \begin{subfigure}[t]{0.32\textwidth}
        \centering
        \includegraphics[width=\columnwidth]{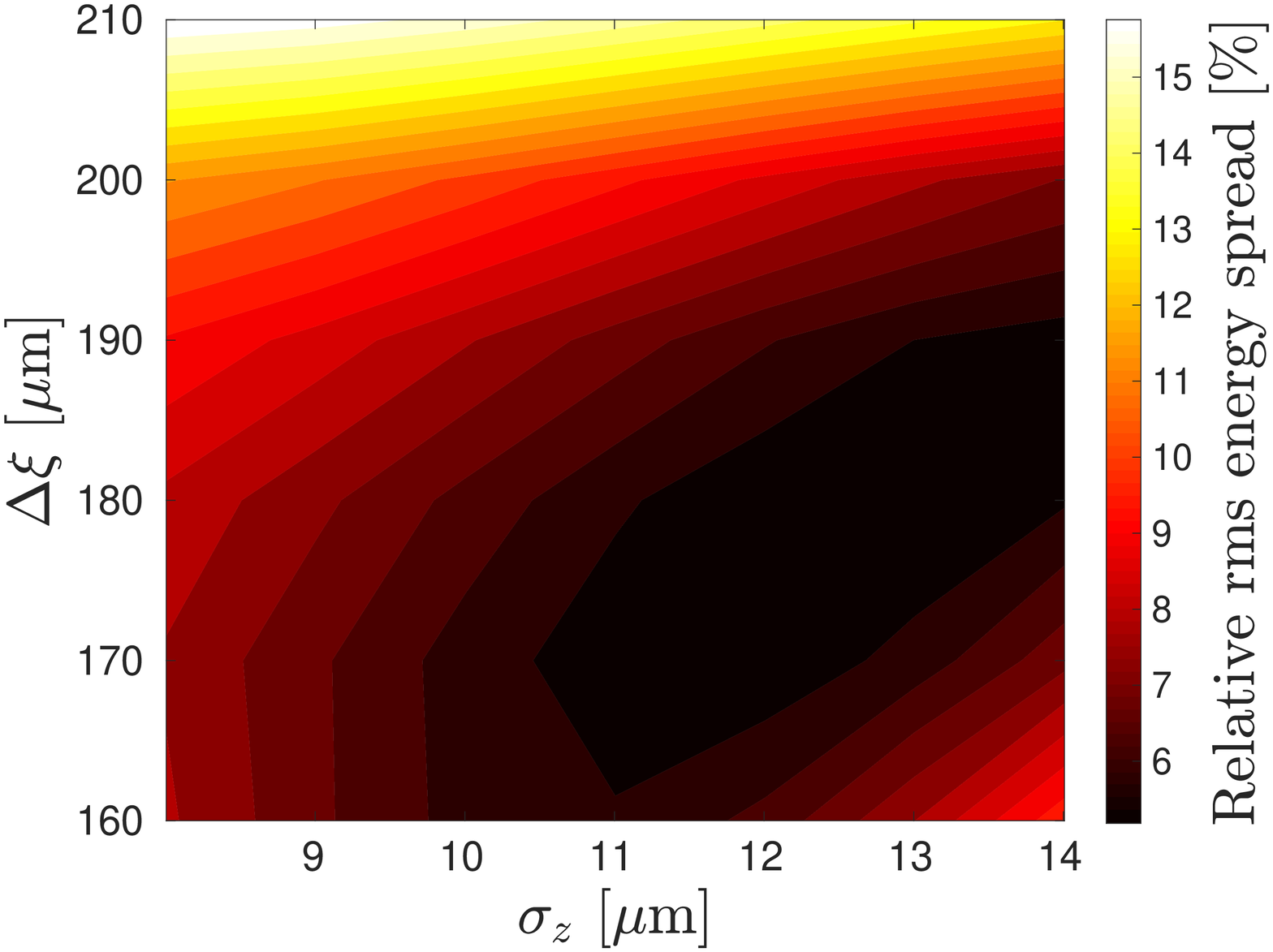}
        \caption{$N_\mathrm{MB}=\SI{e10}{}$.}
        \label{2019-07-19_sigmaE_E_inf_contourPlot_N1e+10}
    \end{subfigure}
    
\caption{Relative rms energy spread vs. beam separation distance $\Delta\xi$ and the rms beam length $\sigma_z$ for main beams with various particle numbers $N_\mathrm{MB}$. The rest of the parameters are from Snowmass parameters.}
\end{figure}

For an initially monochromatic beam with $N$ electrons divided into $n$ slices, the variance of energy is given by
\begin{equation}
    \sigma_\calE^2 = \frac{1}{N}\sum_{i=1}^n N_i \left( \calE_i-\mean{\calE} \right)^2 = \frac{1}{N}\sum_{i=1}^n N_i\left( -eE_{\parallel i} s + e\mean{E_\parallel}s \right)^2,
\end{equation}
where $N_i$ and $E_{\parallel i}$ are the number of electrons and the total longitudinal field acting on beam slice $i$ respectively. The relative rms energy spread is then given by
\begin{equation}
    \frac{\sigma_\calE}{\mean{\calE}} = \frac{es}{\mean{\calE_0}-e\mean{E_\parallel}s}\sqrt{ \frac{1}{N} \sum_{i=1}^n N_i \left( \mean{E_\parallel}-E_{\parallel i} \right)^2},
\label{eq:sigmaE_E_(s)}
\end{equation}
where $\mean{\calE_0}$ is the mean initial energy. In the limit $s\to\infty$, this reduces to
\begin{equation}
    \frac{\sigma_\calE}{\mean{\calE}} = -\frac{1}{\mean{E_\parallel}}\sqrt{ \frac{1}{N} \sum_{i=1}^n N_i \left( \mean{E_\parallel}-E_{\parallel i} \right)^2}.
\label{eq:sigmaE_E_Limit}
\end{equation}
Thus, by using equation \eqref{eq:sigmaE_E_Limit}, the final energy spread can be extrapolated from the initial longitudinal field $E_\parallel(\xi)$, again assuming that $E_\parallel(\xi)$ and the longitudinal particle number distribution do not change significantly during propagation. By extracting $E_\parallel(\xi)$ from QuickPIC simulation results using various combinations of main beam particle number $N_\mathrm{MB}$, rms main beam beam length $\sigma_z$ and beam separation distance $\Delta\xi$, we obtained a series of contour plots for ${\SI{2e9}{}\leq N_\mathrm{MB} \leq\SI{e10}{}}$ that provide an overview over the effect of $N_\mathrm{MB}$, $\sigma_z$ and $\Delta\xi$ on the energy spread. Three examples of such contour plots are shown in figure \ref{fig:2019-07-19_sigmaE_E_inf_contourPlot_N8e+09}-\ref{2019-07-19_sigmaE_E_inf_contourPlot_N1e+10}. Such an overview reveals the region of minimum energy spread in the $\sigma_z$-$\Delta\xi$ plane for various charges, which is crucial for the study of accelerator parameters. Such contour plots are however limited by the chosen resolution of the simulations, so that the distance between actual data points in the $\sigma_z$-direction is \SI{1}{\micro\meter}, and \SI{10}{\micro\meter} in the $\Delta\xi$-direction. This applies to all contour plots in this paper. Furthermore, due to the simulation resolution, $\sigma_z$ was chosen to be $\geq\SI{2}{\micro\meter}$.








The normalized amplitude \cite{normalizedAmplitude} defined as
\begin{equation}
    \Lambda(s) =\sum\limits_{i=1}^n(X_{\sub{N}i}(s)^2 + X_{\sub{N}i}'(s)^2) = \sum\limits_{i=1}^n\left[ \left( \frac{X_i(s)}{\sigma_x(s)} \right)^2 + \left( \frac{X_i'(s)}{\sigma_{x'}(s)} \right)^2 \right],
\end{equation}
where
\begin{equation}
    \sigma_x(s) = \sqrt{\frac{\beta(s)\varepsilon_{\mathrm{N}x}}{\gamma(s)}}, \quad \sigma_{x'}(s) = \sqrt{\frac{\varepsilon_{\mathrm{N}x}}{\gamma(s)\beta(s)}}
\end{equation}
and $\varepsilon_{\mathrm{N}x}$ is the normalized emittance, remains constant in the absence of transverse wakefields. The normalized amplification factor $\Lambda_\mathrm{final}/\Lambda_\mathrm{initial}$ can thus be used to quantify the amplification of the transverse jitter of the incoming beam.

For a main beam with charge $Q_\mathrm{MB}$ accelerated in the the wake excited by a drive beam with charge $Q_\mathrm{DB}$, the drive beam to main beam efficiency is defined as
\begin{equation}
    \eta = \frac{\Delta\calE_\mathrm{MB}}{\calE_\mathrm{DB}}\frac{Q_\mathrm{MB}}{Q_\mathrm{DB}},
\end{equation}
where $\Delta\calE_\mathrm{MB}$ is the energy gain of the main beam, $\calE_\mathrm{DB}$ is the initial drive beam energy. This definition considers all the energy of the DB as spent regardless of how much energy has been extracted. Assuming the drive beam's energy is fully depleted in a plasma of length $L_\mathrm{d}$, the efficiency can also be written as
\begin{equation}
    \eta=\frac{E_\mathrm{A}L_\mathrm{d}}{E_\mathrm{D}L_\mathrm{d}}\frac{Q_\mathrm{MB}}{Q_\mathrm{DB}}=T\frac{Q_\mathrm{MB}}{Q_\mathrm{DB}},
\label{eq:efficiency}
\end{equation}
where $E_\mathrm{D}$ is the peak decelerating field of the drive beam and $E_\mathrm{A}$ is the mean accelerating field of the main beam and $T=E_\mathrm{A}/E_\mathrm{D}$ is the transformer ratio.

\subsection{Results}



Using the developed framework, the relation between energy spread, instability and efficiency can now be studied. A main beam parameter scan over ${\SI{2e9}{}\leq N_\mathrm{MB} \leq\SI{e10}{}}$, $\sigma_z$ and $\Delta\xi$ using the Snowmass $T=1$ parameter set as a basis has been performed to obtain values for $\sigma_\calE/\mean{\calE}$, $\Lambda/\Lambda_0$ and $\eta$. $\sigma_\calE/\mean{\calE}$ and $\eta$ were calculated from a single QuickPIC time step using equation \eqref{eq:sigmaE_E_Limit} and \eqref{eq:efficiency} respectively, whereas $\Lambda/\Lambda_0$ for a $\SI{1.5}{\tera\electronvolt}$ accelerator was calculated using the simplified quasi-static model. We initially assumed that all parameters used in the parameter scan were able to generate main beams sufficiently stable to satisfy the basic assumptions for the simplified quasi-static model. Large values of $\Lambda/\Lambda_0$ should therefore be ignored in the end results.

\begin{figure}[ht]
\begin{minipage}[t]{16pc}

    \centering
    \includegraphics[width=\columnwidth]{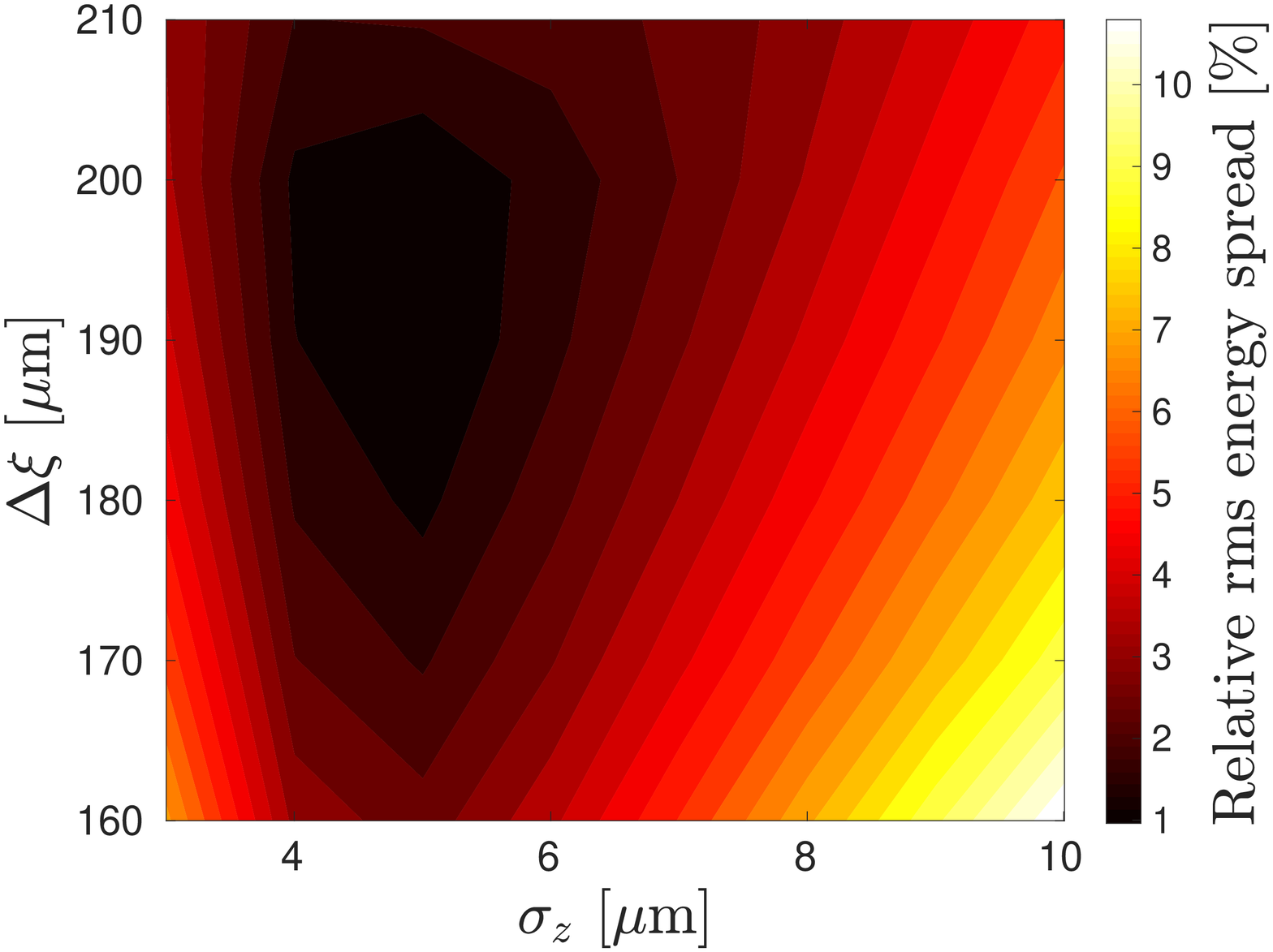}
    \caption{Contour plot of the relative rms energy spread in the $\sigma_z$-$\Delta\xi$ plane for a main beam with $\SI{5e9}{}$ electrons .}
    \label{fig:contourPlot_sigmaEE}

\end{minipage}\hspace{1pc}%
\begin{minipage}[t]{16pc}

    \centering
    \includegraphics[width=\columnwidth]{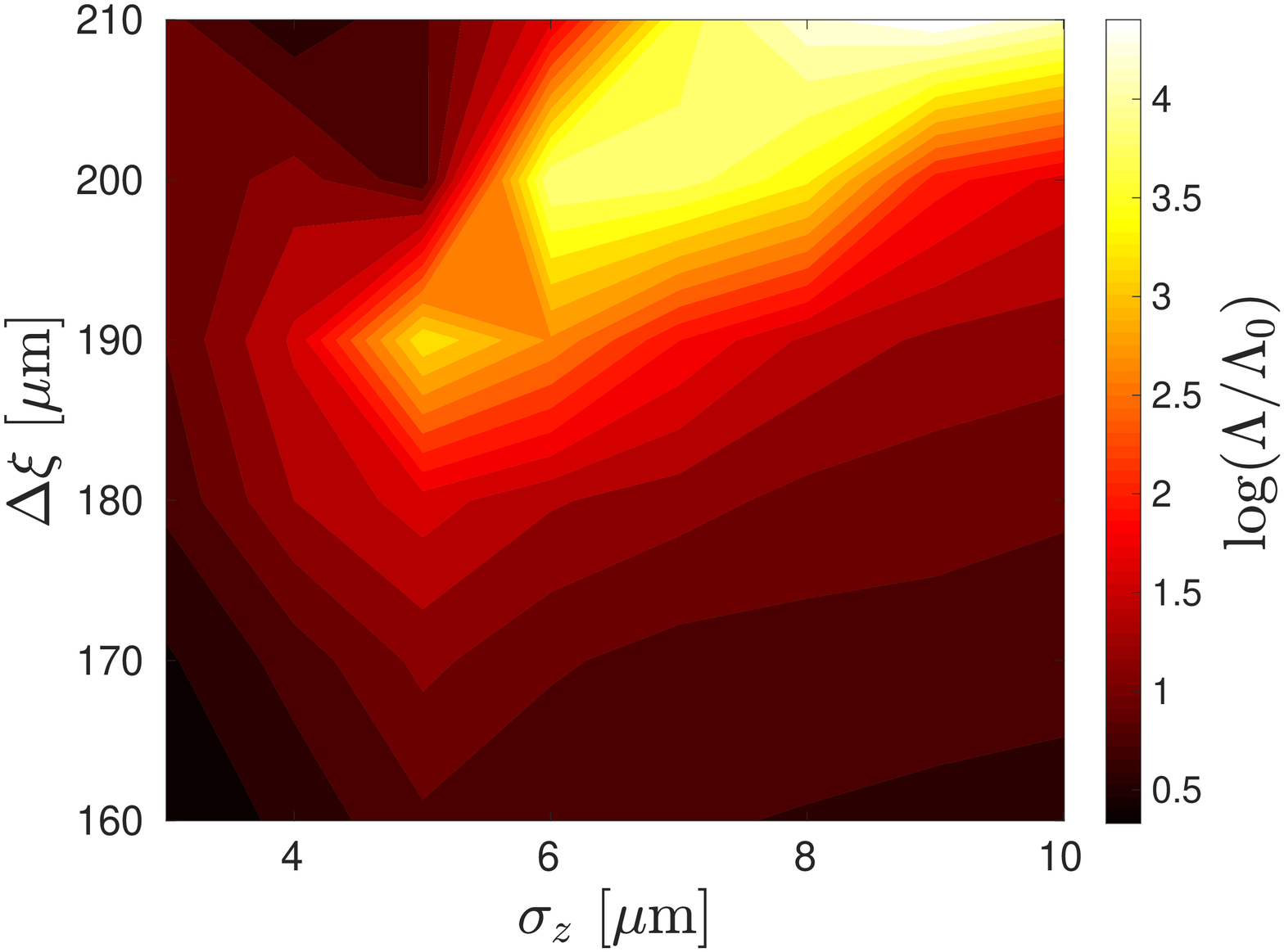}
    \caption{Contour plot of the normalized amplification factor in the $\sigma_z$-$\Delta\xi$ plane for a main beam with $\SI{5e9}{}$ electrons.}
    \label{fig:contourPlot_normAmp}
    
\end{minipage}\hspace{1pc}
\end{figure}

\begin{figure}[h]
\begin{minipage}[t]{16pc}

    \centering
    \includegraphics[width=\columnwidth]{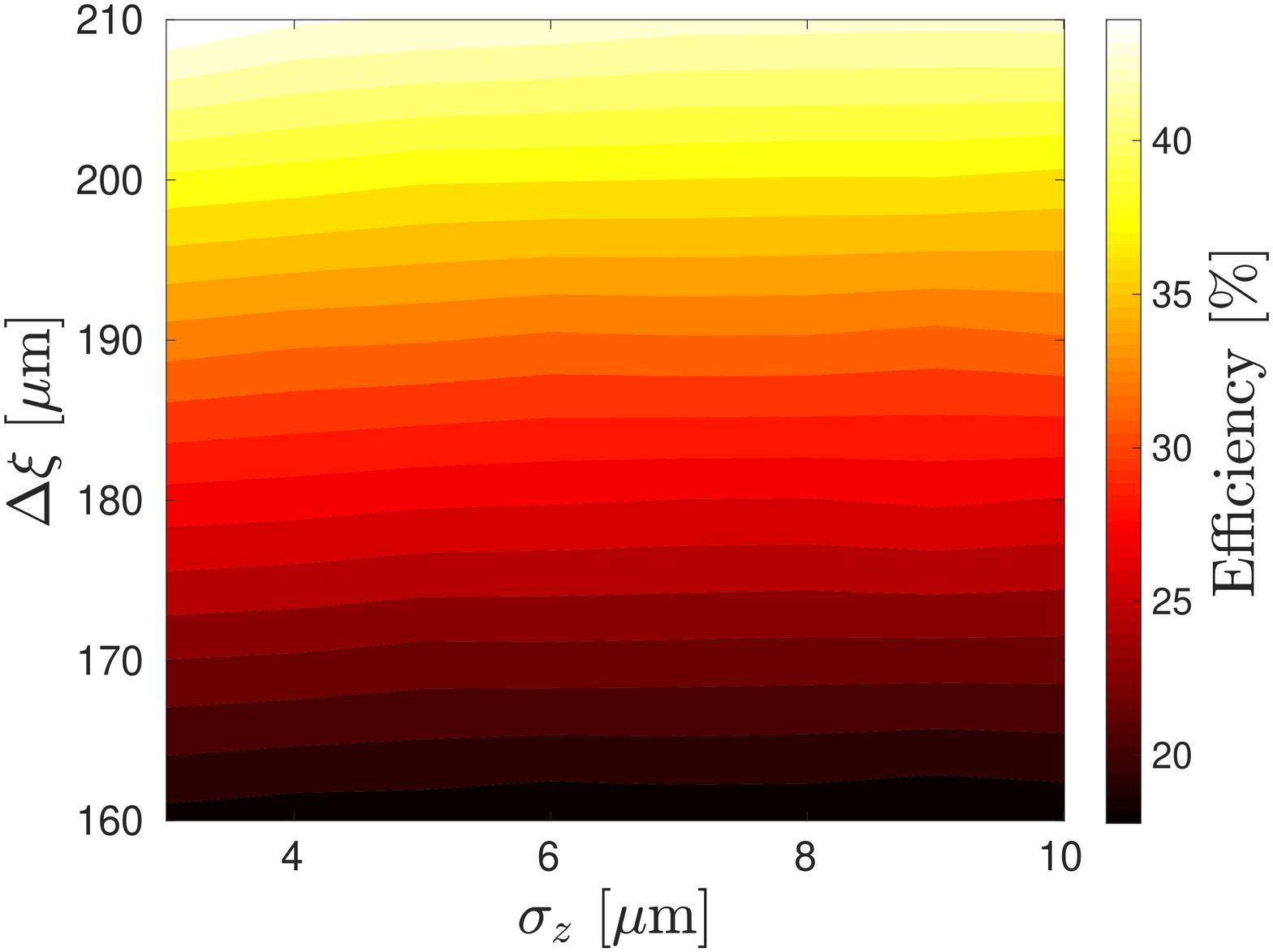}
    \caption{Contour plot of the efficiency in the $\sigma_z$-$\Delta\xi$ plane for a main beam with $\SI{5e9}{}$ electrons.}
    \label{fig:contourPlot_efficiency}

\end{minipage}\hspace{1pc}%
\begin{minipage}[t]{16pc}

    \centering
    \includegraphics[width=\columnwidth]{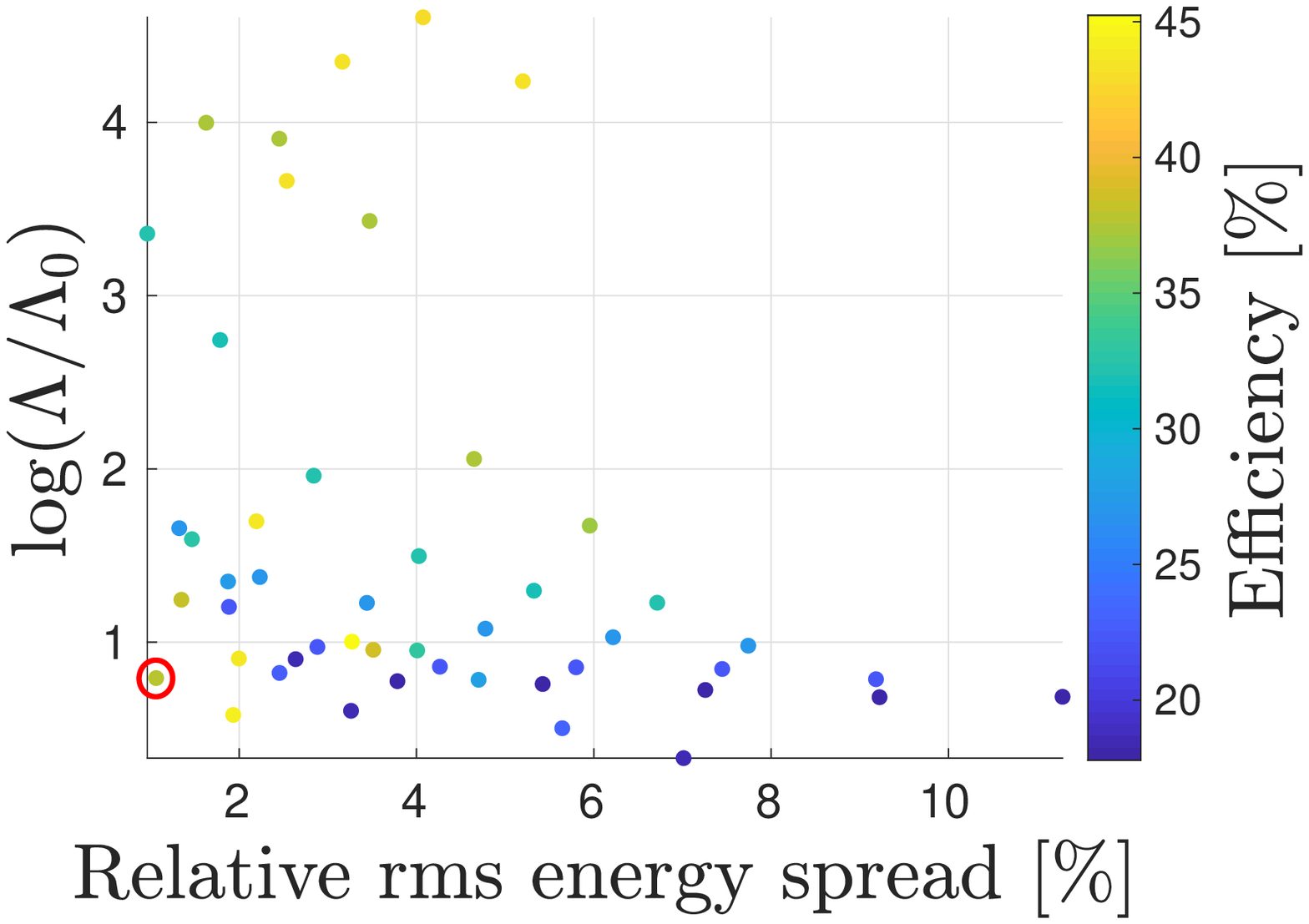}
    \caption{Relation between relative rms energy spread, normalized amplification factor and efficiency for a main beam with $\SI{5e9}{}$ electrons. A potential candidate for a new parameter set is marked with a red circle.}
    \label{fig:scatterPlot_sigmaEE_normAmpFac_efficiency}
    
\end{minipage}\hspace{1pc}
\end{figure}

The results for $N_\mathrm{MB}=\SI{5e9}{}$ are shown in figure \ref{fig:contourPlot_efficiency}-\ref{fig:scatterPlot_sigmaEE_normAmpFac_efficiency}, as this particle number resulted in the most desirable results. Figure \ref{fig:contourPlot_sigmaEE}-\ref{fig:contourPlot_efficiency} can be used to identify regions in the $\sigma_z$-$\Delta\xi$ plane containing desirable values for $\sigma_\calE/\mean{\calE}$, $\Lambda/\Lambda_0$ and $\eta$. An optimal set of parameters for an accelerator would however require low values for $\sigma_\calE/\mean{\calE}$ and $\Lambda/\Lambda_0$, while high values of $\eta$ are desirable. These requirements can be conflicting, so in order to arrive at a reasonable compromise, the data is combined to obtain an overview shown in figure \ref{fig:scatterPlot_sigmaEE_normAmpFac_efficiency}.

\begin{figure}[h]
        \includegraphics[width=15pc]{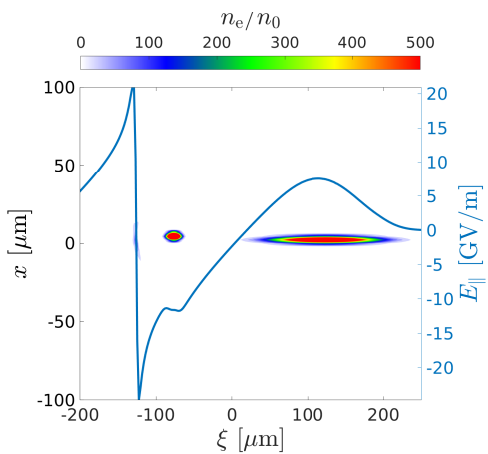}\hspace{2pc}%
        \begin{minipage}[b]{14pc}\caption{Initial electron number density $n_\mathrm{e}$ per unit initial plasma density $n_0$ and the total longitudinal electric field $E_\parallel(\xi)$ obtained from a QuickPIC simulation with the new parameters.}
        \label{fig:QEB+QEP_optimal}
    \end{minipage}
\end{figure}

Figure \ref{fig:scatterPlot_sigmaEE_normAmpFac_efficiency} shows several data points of interest, for instance the point $\sigma_\calE/\mean{\calE}=1.1\%$, $\log(\Lambda/\Lambda_0)=0.8$, $\eta=37.5\%$ marked with a red circle, which corresponds to $\sigma_z=\SI{5}{\micro\meter}$, $\Delta\xi=\SI{200}{\micro\meter}$. A corresponding plot for the initial electron number density and the longitudinal field obtained from QuickPIC simulation is shown in figure \ref{fig:QEB+QEP_optimal}. 

These main beam parameters provide improvements over the Snowmass parameter set both in terms of energy spread and stability, but result in a lower efficiency. These parameters and results are summarized in table \ref{tab:parameterComparison}, where the energy spread for the Snowmass parameter set has been re-calculated using the definition \eqref{eq:sigmaE_E_Limit}.

\begin{table}
\caption{Comparison of the Snowmass $T=1$ parameter set and the new parameter set.}
\label{tab:parameterComparison}
    \begin{center}
    \begin{tabular}{lll}
    \br
                    &   Snowmass    &   New parameters\\
    \mr
    $N_\mathrm{MB}$ [\SI{e9}{}]             &   10  &   5\\
    $\sigma_z$ $[\SI{}{\micro\meter}]$     &   20   &   5\\
    $\Delta\xi$ $[\SI{}{\micro\meter}]$     &   187   &   200\\
    $\sigma_\calE/\mean{\calE}$ $[\%]$     &   12   &   1.1\\
    $\Lambda/\Lambda_0$                     &   \SI{6.7e2}{}  &   6\\
    $\eta$ $[\%]$                           &   50   &   37.5\\
    \br
\end{tabular}
\end{center}
\end{table}

A PWFA multi-\SI{}{\tera\electronvolt} accelerator can be envisioned to be used as the main linac for a linear collider. A core metric of performance for a linear collider is luminosity per power, which scales as $\mathcal{L}/P_\mathrm{AC} \propto \eta/\sqrt{\sigma_z}$
when beam strahlung has been taken into account, assuming that the horizontal beam size can be made sufficiently small, and that the vertical beam size is kept constant \cite{CLIC_CDR}. Assuming that this can be achieved for the new parameter set, the luminosity per power is actually 1.5 times higher than the corresponding value provided by the Snowmass parameters, even though the new parameter set offers a lower efficiency.

\section{Conclusion}
Even though several conceptual parameter sets for a PWFA-LC have been proposed, no PWFA-LC studies have so far considered the constraint of efficiency imposed by transverse instabilities. 

In this paper, we described the transverse instabilities in PWFA using the wakefield formalism and benchmarked the results against QuickPIC simulation results. Using the wakefield formalism, a simplified quasi-static model was developed, and was combined with QuickPIC simulations in order to model the evolution of the transverse oscillations of the main beam over a $\SI{1.5}{\tera\electronvolt}$ PWFA accelerator. 

We demonstrated that the Snowmass parameter set was unable to provide stable propagation for a main beam consisting of electrons, and we performed a parameter scan over the main beam charge, rms beam length and beam separation distance using the Snowmass parameter set as a basis. The parameter scan provided a new set of parameters that improved the Snowmass parameter set in terms of energy spread, stability and luminosity per power. This parameter study for the main electron beam is however not exhaustive, and did not consider the effects of beam induced ion motion, which has been shown to mitigate hosing \cite{IonMotion}. Furthermore, tolerance studies still remain to be performed, which will be included in future works.

\section*{References}
\bibliography{references}
\end{document}